# Photometric Investigation of WY Leo: An Eclipsing Binary System with δ Scuti Component

Filiz Kahraman Aliçavuş[1]

**Abstract-** Eclipsing binary stars with δ Scuti components are exceptional systems for gaining a deeper understanding of stellar systems. WY Leo is one such system exhibiting δ Scuti-type oscillations. However, it has not been extensively studied in the literature. WY Leo was observed by the Transiting Exoplanet Survey Satellite (TESS), providing two sectors of high-quality photometric data. This study focused on the photometric analysis of WY Leo. Binary modeling was performed, and the system's fundamental stellar parameters, such as mass and radius, were determined. The pulsational properties of WY Leo were also investigated, revealing that the more luminous star exhibits δ Scuti-type oscillations with a period of 0.052 days. The position of the primary component was examined on the HR diagram and found to lie within the δ Scuti instability strip.

## 1. Introduction

The importance of eclipsing binary systems has been known for decades. These systems are unique tools to estimate the fundamental stellar parameters such as mass (*M*) and radius (*R*) with a reasonable accuracy [1, 2]. With the combination of high-quality photometric and spectroscopic data, the fundamental stellar parameters could be estimated with more precision. The importance of eclipsing binary systems increased with the discovery that one of their components pulsates [3, 4].

Pulsating stars exhibit radial and/or non-radial oscillations with varying pulsation periods and span a range of spectral types from approximately B to G [5]. One of the most well-known groups of pulsating stars is the δ Scuti variables, which were first discovered in eclipsing binaries [3, 4]. δ Scuti variables are A-F type stars with oscillation periods ranging from about 20 minutes to 8 hours [5]. These pulsations can be used to probe the interior structure of stars [6–8], making oscillating stars highly significant. Consequently, eclipsing binaries that exhibit oscillations provide valuable insights for understanding stellar systems in greater detail.

In a recent study, [9] investigated eclipsing binaries with δ Scuti pulsators, listing around 90 such systems along with their properties, including *M*, *R*, pulsation period, and amplitude. The number of known δ Scuti stars in eclipsing binaries has grown with the analysis of space-based data [10–12], leading to the discovery of many new systems. To better understand the nature and pulsational

[1]filizkahraman01@gmail.com (Corresponding Author)
[1]Department of Physics, Faculty of Science, Çanakkale Onsekiz Mart University, Çanakkale, Türkiye
[1]Astrophysics Research Center and Ulupınar Observatory, Çanakkale Onsekiz Mart University, Çanakkale, Türkiye

structure of δ Scuti stars in eclipsing binaries, these systems are essential objects of study. Therefore, the growing number of such systems is crucial. In this study, WY Leo analyzes one δ Scuti star in an eclipsing binary.

WY Leo (V = 10.89 mag, TIC 192984258) is a semidetached eclipsing binary system with a spectral type of A2+K2IV, as reported by [13]. In a recent study based on LAMOST spectroscopy, the system's new spectral type was identified as kA6hA9mF0V, with an effective temperature of 7408 K. The system's orbital period is 4.985854 days, according to [14]. However, there has been no detailed system analysis in the literature. The pulsational nature of WY Leo was only briefly investigated by [15], who classified the system as an eclipsing binary with a δ Scuti component. Therefore, this study considered high-quality space-based data to conduct a detailed photometric analysis. The paper is organized as follows: in Section 2, details about the photometric data are provided; in Sections 3 and 4, the binary modeling and pulsational analysis of the system are given; and in Section 5, the discussion and conclusions are presented.

## 2. Photometric Data

The study utilized the high-resolution light curve of WY Leo from the Transiting Exoplanet Survey Satellite (TESS) [16] to enhance the precision of the fundamental parameters. TESS was designed to observe exoplanets around bright, close stars (with a TESS magnitude < 13.5). The space telescope observes the sky by dividing it into sectors. TESS performs approximately 27 days of observation within each sector, with varying exposure times of 120, 200, 600, and 1800 seconds. TESS data are available through the Barbara A. Mikulski Archive for Space Telescopes (MAST) portal (https://mast.stsci.edu) and include fluxes from both simple aperture photometry (SAP) and pre-search data conditioning SAP (PDCSAP).

WY Leo was observed in two sectors, 46 and 72, with exposure times of 600 and 200 seconds, respectively. Data from both sectors were incorporated into the study, and SAP fluxes were used for the analysis. These fluxes were converted into magnitudes using the equation provided by [10].

The photometric TESS data were applied to binary modeling and to investigate the pulsational structure of the system. Since WY Leo exhibits δ Scuti type pulsations, as identified by [15], and δ Scuti tars typically show oscillations in the frequency range of ∼ 5-80 $d^{-1}$ [17], the 200-second TESS data were used for pulsational analysis, owing to its Nyquist frequency of ∼ 215 $d^{-1}$. The 600-second TESS data were used for binary modeling.

## 3. Binary Modeling

The 600-second TESS data were used for binary modeling of WY Leo. Since WY Leo exhibits significant oscillations that affect the eclipse profiles, the pulsational variation was first removed from the entire light curve. To achieve this, the PERIOD04 program [18] was employed. The binary variation of the system was modeled, considering its orbital period ($P_{orb}$) (taken from [14] as 4.98591 ± 0.00001) and the harmonics of $P_{orb}$. Then, the pulsations were removed from the entire light curve using the same method as in [12]. The oscillation-removed light curve is shown in Figure 1.

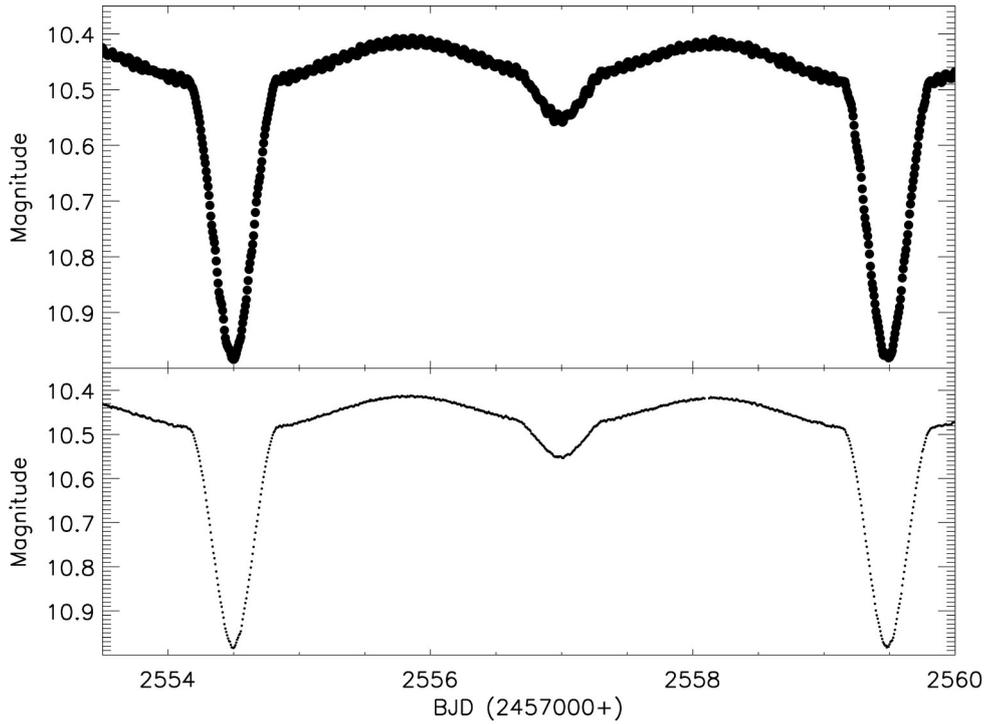

**Figure 1.** Upper panel: Original binary light curve of WY Leo. Lower panel: pulsation-removed light curve

The oscillation-removed light curve was subsequently analyzed using the Wilson-Devinney binary modeling code [19], combined with Monte Carlo simulations [20, 21]. In this analysis, some parameters were fixed, such as the effective temperature ($T_{eff}$) of the more luminous star (primary star), and also the bolometric albedos [22], the bolometric gravity-darkening coefficient [23], the logarithmic limb darkening coefficient [24] as mentioned in [25]. The $T_{eff}$ value of the primary stars was taken as 7408 ± 40 K from [26]. They estimated this value based on the analysis of the low-resolution LAMOST spectrum. This $T_{eff}$ value was considered because during the analysis of LAMOST data, mostly the blue part of the spectrum was taken into account, where the flux of the cool secondary component is negligibly low, and the spectrum primarily represents the primary component. Therefore, this $T_{eff}$ value was adopted during the analysis. Additionally, in the analysis, the following parameters were searched for: the mass ratio ($q$), dimensionless potential ($\Omega$), the $T_{eff}$ of the secondary component, inclination ($i$), phase shift ($\phi$), the relative radius ($r$) and third body light contribution ($l_3$). The semi-detached binary configuration was considered in the analysis, as it was classified as a semi-detached system by [13].

Since WY Leo lacks radial velocity analysis, one of the most important parameters, $q$, is unknown. Therefore, a classical approach was used, involving a search for the $q$ value before finalizing the binary modeling. The $q$ search was initially conducted with a step size of 0.1 within the range of 0.1 to 1, and the minimum sum of squared residuals ($\Sigma(O-C)^2$) was identified. After locating the minimum $\Sigma(O-C)^2$ value, the $q$ search was refined with a step size of 0.001 around the $q$ value that produced the minimum $\Sigma(O-C)^2$ from the initial search. As a result, the $q$ value was determined to be 0.285 ± 0.090. The uncertainty in the $q$ value was estimated at the 1$\sigma$ level.

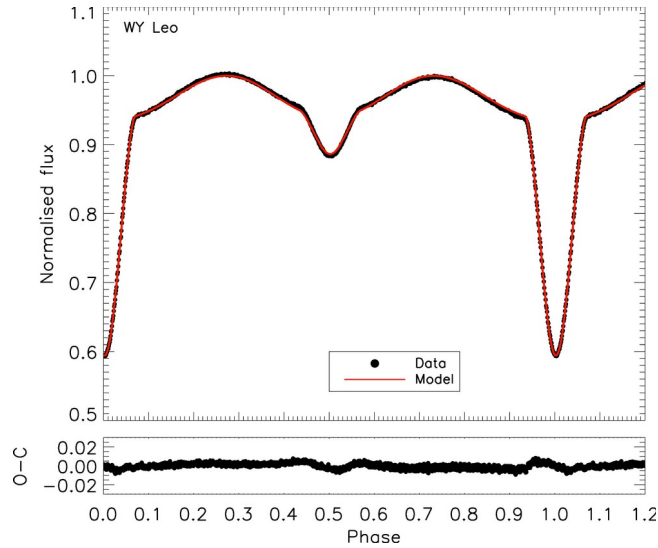

**Figure 2.** Upper panel: Theoretical binary modeling (red line) of WY Leo. Lower panel: residuals

After finding the $q$ value, the final binary modeling was done by setting $q$ as fixed. As a result of the analysis, the binarity parameters of WY Leo were found. The resulting parameters and the fit of the theoretical binary model to the observations are shown in Table 1 and Figure 2, respectively. Utilizing the final parameters determined from the binary modeling, the fundamental stellar parameters such as $M$, $R$, bolometric magnitude ($M_{bolo}$), absolute magnitude ($M_V$) in addition to surface gravity value (log $g$) were calculated. For these calculations, the bolometric corrections were taken from [27] and the interstellar extinction ($A_V$) was estimated to be 0.077 mag from [28]. The $M$ of the primary component was estimated using the $T_{eff}$ value and the study by [29]. The uncertainty in the primary component's mass was calculated using the error in $T_{eff}$. The $M$ of the secondary component was determined using the mass ratio ($q$). The semi-major axis ($a$) was then calculated based on the determined masses and Kepler's third law, yielding a value of 16.34 $R_⊙$. Finally, the $R$ values of both binary components were derived using the values of $a$ and $r$. The system's distance was also determined using the distance modulus equation. All estimated results are summarized in Table 1.

**Table 1.** Results of the binary modeling and the fundamental stellar parameters

| Parameter | Value | Parameter | Value |
|---|---:|---|---:|
| $i$ (°) | 75.88 ± 0.01 | $M_1$ ($M_⊙$) | 1.85 ± 0.05 |
| $T_{eff1}$* (K) | 7408 ± 40 | $M_2$ ($M_⊙$) | 0.52 ± 0.05 |
| $T_{eff2}$ (K) | 4410 ± 240 | $R_1$ ($R_⊙$) | 3.19 ± 0.05 |
| $Ω_1$ | 5.409 ± 0.008 | $R_2$ ($R_⊙$) | 4.36 ± 0.07 |
| $Ω_2$ | 2.433 ± 0.009 | log ($L_1/L_⊙$) | 1.44 ± 0.03 |
| Phase shift | 0.0024 ± 0.0001 | log ($L_2/L_⊙$) | 0.81 ± 0.02 |
| $q$* | 0.285 ± 0.090 | log $g_1$ (cgs) | 3.70 ± 0.03 |
| $r_1$ (mean) | 0.1959 ± 0.0003 | log $g_2$ (cgs) | 2.87 ± 0.07 |
| $r_2$ (mean) | 0.2680 ± 0.0008 | $M_{bolo1}$ (mag) | 1.14 ± 0.02 |
| $L_1 / (L_1+L_2)$ | 0.74 ± 0.01 | $M_{bolo2}$ (mag) | 2.72 ± 0.04 |
| $L_2 / (L_1+L_2)$ | 0.26 ± 0.01 | $M_{V1}$ (mag) | 1.07 ± 0.06 |
| $l_3$ | 0.0 | $M_{V2}$ (mag) | 3.33 ± 0.08 |
|  |  | Distance (pc) | 985 ± 32 |

Subscripts 1-3 represent the primary, secondary, and third binary components, respectively.
$^h$ shows the fixed parameters

## 4. Pulsating Analysis

After modeling the binary variation of WY Leo, the pulsation variation was analyzed using the 200-second TESS data, which had been cleaned of binary variations. The system's oscillations were investigated using the PERIOD04 program, which determines pulsational frequencies based on the discrete Fourier transform algorithm. A significant threshold of 4.5σ was applied for the determined frequencies by the study by [30]. The analysis revealed seven significant frequencies, with the highest amplitude frequency of 19.15 $d^{-1}$ and an amplitude of 2.02 mmag. A list of the determined frequencies is provided in Table 2 values of the binary provides a list of the determined frequencies components; the pulsating component of the WY Leo system was identified as the hotter primary star. The oscillating star exhibits δ Scuti type variation, as indicated by its pulsation frequencies. The pulsation amplitude spectrum and the theoretical pulsation fit to the observations are shown in Figure 3.

**Table 2.** Resulting frequencies, amplitudes, and the phases for the significant frequencies

|  | Frequency ($d^{-1}$) | Amplitude (mmag) | Phase (rad) | S/N |
|---|---|---|---|---|
| $\nu_1$ | 19.1489(2) | 2.02 | 0.764(2) | 15 |
| $\nu_2$ | 18.4245(2) | 1.68 | 0.642(3) | 11 |
| $\nu_3$ | 18.9127(2) | 1.10 | 0.785(3) | 8 |
| $\nu_4$ | 18.3673(3) | 1.08 | 0.967(4) | 7 |
| $\nu_5$ | 14.9360(3) | 0.76 | 0.605(4) | 9 |
| $\nu_6$ | 22.7239(3) | 0.75 | 0.146(4) | 9 |
| $\nu_7$ | 18.1193(3) | 0.66 | 0.302(4) | 5 |

The pulsation constant for the pulsating primary component was calculated based on the equation of $Q = P_{puls} \sqrt{\rho/\rho_\odot} = P_{puls} M^{1/2} R^{-3/2}$. The $M$ and $R$ values were taken from the results of the binary modeling, and the $P_{puls}$ is the periods of the pulsation frequency. The pulsation constants were calculated between 0.011 and 0.016 d, which shows that the frequencies are the 3rd to 6th harmonics of the fundamental modes according to the study of [31].

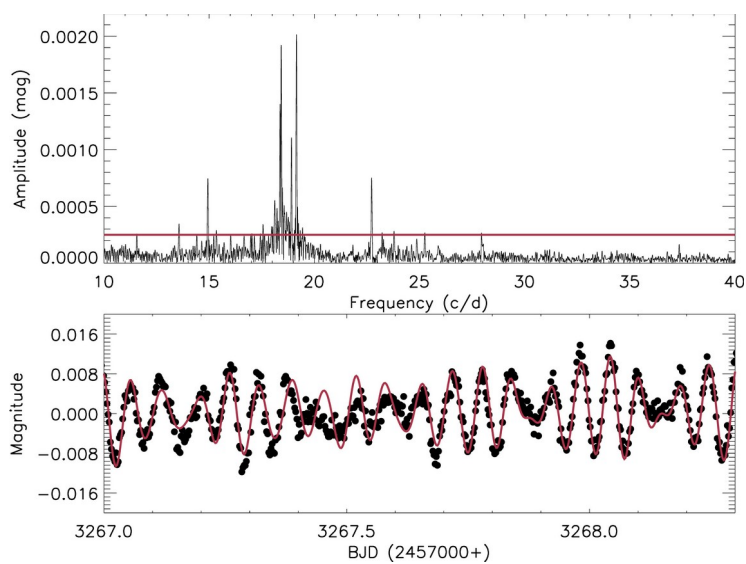

**Figure 3.** Upper panel: Amplitude spectrum of the pulsating primary component of WY Leo. The red horizontal line shows a 4.5σ level. Lower panel: Theoretical pulsation fits the observation. The red line in this panel represents the theoretical pulsation fit.

The position of the pulsating primary component on the Hertzsprung-Russell (HR) diagram was also examined. For this purpose, the observational instability strip for δ Scuti stars was taken from [32], and the evolutionary models were obtained from the MESA Isochrones and Stellar Tracks (MIST) [33]. It is important to note that these MIST models are for single stars and do not account for binary effects. The position of the pulsating primary component on the HR diagram is shown in Figure 4. As can be seen from the figure, the pulsating component lies just inside the δ Scuti instability strip.

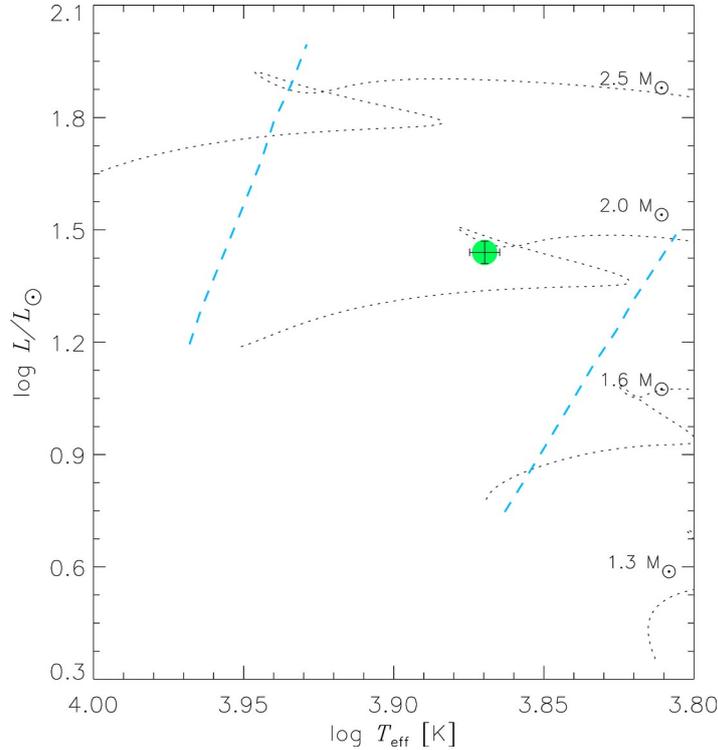

**Figure 4.** The position of the pulsating component on the HR diagram. The dashed blue and dotted black lines represent the δ Scuti stability strip and evolutionary tracks, respectively.

## 5. Conclusion

This study presents the WY Leo system's first detailed photometric analysis using high-quality TESS data. The binary system's parameters were determined by modeling its light curve variation. Based on this analysis, the masses of the binary components were found to be 1.85(5) and 0.52(2) $M_\odot$ for the primary and secondary stars, respectively. Based on the binary modeling results, The system's distance was estimated to be 985 ± 32 pc. This value was close to Gaia distance ∼ 940 ± 22 pc within the error margins. In addition to the mass values, the components' $R$ and $L$ parameters were calculated, and the position of the primary component on the HR diagram was determined. The primary star was found to lie within the δ Scuti instability strip. A pulsational analysis revealed that the primary, more luminous star exhibits δ Scuti type oscillations. The highest amplitude pulsation frequency was identified as 19.1489 $d^{-1}$, which differs from the 15.2428 $d^{-1}$ value reported in the first literature study [15]. Although our study also detected this frequency, it appears to be a harmonic of the orbital period's frequency. Therefore, it was excluded from further analysis.

For the δ Scuti stars in the eclipsing binaries, there are well-known relationships between the pulsation period, amplitude, and some binary parameters such as orbital period-pulsation period ($P_{orb}$ - $P_{puls}$) relationship [9, 34]. According to the updated pulsational parameters for the pulsation component of WY Leo, the new position of the system on the logP$_{orb}$ - logP$_{puls}$ relationship was examined and is shown in Figure 5. As seen from the figure, the updated position of the system slightly differs.

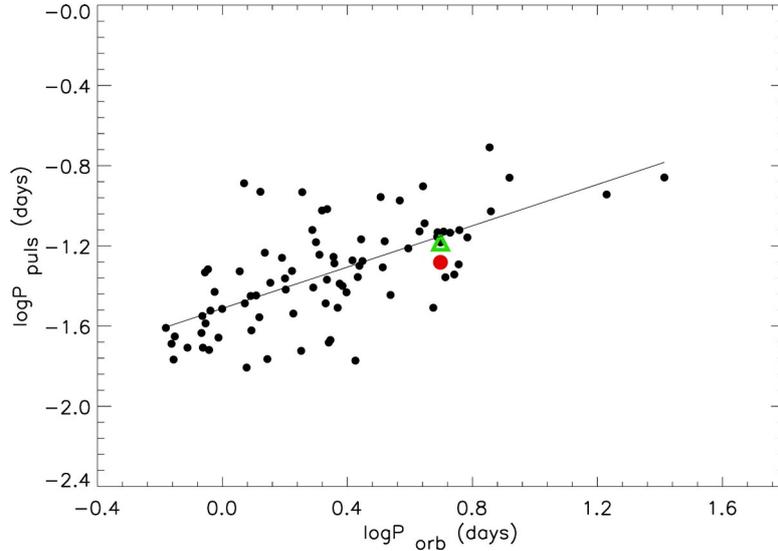

**Figure 5.** The position of the pulsating component on the logP$_{orb}$ - logP$_{puls}$ relationship. The green triangle and red dot represent the position based on the literature data [15] and the updated position, respectively. The black dots are the eclipsing binaries containing δ Scuti components, and taken from [9]

Another well-known relationship given for δ Scuti stars in eclipsing binaries is the log $g$ - logP$_{puls}$ relationship. The position of the primary component of WY Leo is shown in this relationship as well in Figure 6. The position of the system matches well with the other eclipsing binaries containing δ Scuti components.

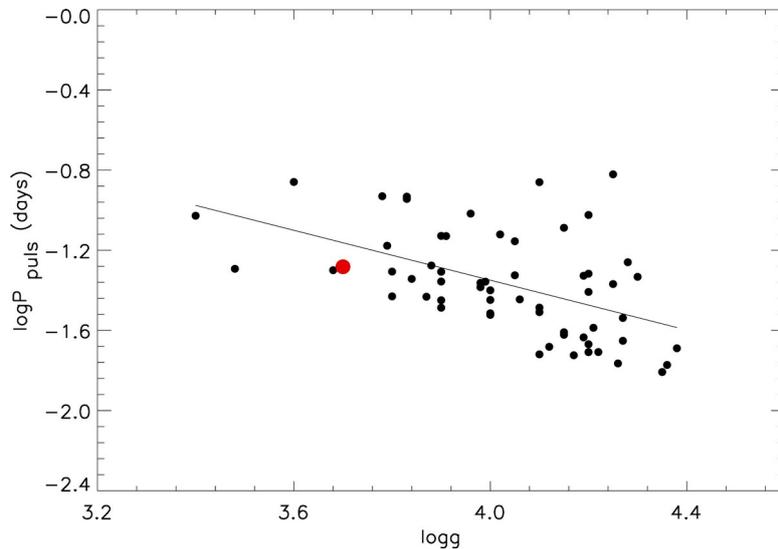

**Figure 6.** The position of the pulsating component on the log $g$-log $P_{puls}$ relationship. The red dot represents the pulsating primary component of WY Leo. The black dots are the eclipsing binaries containing δ Scuti components, and taken from [9]

As a result, the nature of WY Leo was revealed. Systems like this are crucial for understanding the structure of δ Scuti stars in eclipsing binaries. Analyses supported by photometric and spectroscopic data will be essential for a deeper understanding of these systems. Especially, the estimation of the radial velocity variation of the system would be useful for determining the accurate mass ratio and the masses of the binary components. Further spectroscopic observations of WY Leo will provide valuable insights, allowing for a more detailed system study.

## Author Contributions

The author read and approved the final version of the paper.

## Conflicts of Interest

The author declares no conflict of interest.

## Ethical Review and Approval

No approval from the Board of Ethics is required.

## Acknowledgment

This study has been supported by the Scientific and Technological Research Council (TUBITAK) project through 120F330. The TESS data presented in this paper were obtained from the Mikulski Archive for Space Telescopes (MAST). Funding for the TESS mission is provided by the NASA Explorer Program. This work has made use of data from the European Space Agency (ESA) mission Gaia (http://www.cosmos.esa.int/gaia), processed by the Gaia Data Processing and Analysis Consortium (DPAC, http://www.cosmos.esa.int/web/gaia/dpac/consortium). Funding for the DPAC has been provided by national institutions, in particular, the institutions participating in the Gaia Multilateral Agreement. This research used the SIMBAD database operated at CDS, Strasbourg, France.